# QUASISTATIC OSCILLATIONS IN SUBWAVELENGTH PARTICLES: CAN ONE OBSERVE ENERGY EIGENSTATES?

E. O. Kamenetskii

Microwave Magnetic Laboratory, Department of Electrical and Computer Engineering
Ben Gurion University of the Negev, Beer Sheva, Israel

November 6, 2018

**Abstract**
In increasing the capabilities of the optical and microwave techniques further into the subwavelength regime, quasistatic resonant structures has attracted considerable interest. Electromagnetic responses of electrostatic (ES) plasmon resonances in optics and magnetostatic (MS) magnon resonances in microwaves give rise to a strong enhancement of local fields near the surfaces of subwavelength particles. In the near-field regions of subwavelength particles one can only measure the electric or the magnetic field with accuracy. Such uncertainty in definition of the electric or magnetic field components raises the question of energy eigenstates of quasistatic oscillations. The energy eigenstate problem can be properly formulated when potential functions, used in the quasistatic-resonance problems, are introduced as scalar wave functions. In this case, one should observe quasistatic-wave retardation effects still staying in frames of the quasistatic description of oscillations in a subwavelength particle. In this paper, we analyze the problem of energy quantization of ES resonances in subwavelength optical metallic structures with plasmon oscillations and MS resonances in subwavelength microwave ferrite particles with magnon oscillations. We show that in a case of MS-potential scalar wave function one can observe quasistatic retardation effects and obtain a proper formulation of the energy eigenstate problem.

## I. INTRODUCTION

The optical responses of metal particles arise from collective oscillations of their conduction electrons. The microwave responses of ferrite particles arise from collective oscillations of their precessing electrons. These electromagnetic responses, considering, respectively, as plasmon and magnon resonances, give rise to a strong enhancement of local fields near the particle surfaces. In increasing the capabilities of the optical and microwave techniques further into the subwavelength regime, subwavelength resonant structures has attracted considerable interest. The ability of metal nano structures to support and concentrate electromagnetic energy to spots much smaller than a wavelength arises from the fact that at plasmon oscillations, optical fields are almost purely electric whereas the magnetic field component is small. From the other side, the ability of ferrite structures to support and concentrate electromagnetic energy to spots much smaller than a wavelength arises from the fact that at magnon oscillations, microwave fields are almost purely magnetic whereas the electric field component is small. Conventionally, we will call plasmon oscillations in optical subwavelength resonant structures as electrostatic (ES) resonances and magnon oscillations in microwave subwavelength resonators as magnetostatic (MS) resonances.

In the case of ES resonances in small metallic samples, one neglects a magnetic displacement current and an electric field is expressed via an electrostatic potential, $\vec{E} = -\vec{\nabla}\phi$ [1]. Analogously, the MS-resonance problem is considered as zero-order approximation of Maxwell's equations when one neglects the electric displacement current and expresses a magnetic field is via a magnetostatic potential, $\vec{H} = -\vec{\nabla}\psi$ [2]. Fundamentally, subwavelength sizes of the particles should eliminate any electromagnetic retardation effects. For an electromagnetic wavelength $\lambda$ and particle of size $a$, quasistatic approximation $2\pi a/\lambda \ll 1$ means the transition to a small phase.

What kind of the time-varying field structure one can expect to see when an electric or magnetic displacements currents are neglected and so the electromagnetic-field symmetry (dual symmetry) of Maxwell equations is broken?

When one neglects a displacement current (magnetic or electric) and considers the scalar-function, $\phi(\vec{r},t)$ or $\psi(\vec{r},t)$, solutions, one becomes faced with important questions, whether there could be the propagation behaviors inherent for the quasistatic wave processes and, if any, what is the nature of these retardation effects. In a case of ES resonances, the Ampere-Maxwell law gives the presence of a curl magnetic field. With this magnetic field, however, one cannot define the power-flow density of propagating electrostatic-resonance waves. Certainly, from a classical electrodynamics point of view [3], one does not have a physical mechanism describing the effects of transformation of a curl magnetic field to a potential electric field. In like manner, one can see that in a case of MS resonances, the Faraday law gives the presence of a curl electric field. With this electric field, one cannot define the power-flow density of propagating magnetostatic-resonance waves since, from a classical electrodynamics point of view, one does not have a physical mechanism describing the effects of transformation of a curl electric field to a potential magnetic field [3].

In the subwavelength resonators with ES or MS oscillations, measurements of optical or microwave energies become uncertain since we can measure with accuracy only, respectively, the electric or magnetic field. Thus, an evident questions arise: Do the modes of ES resonances actually diagonalize the total energy of a metal subwavelength particle and, similarly, do the modes of MS resonances actually diagonalize the total energy of the subwavelength ferrite particle? In this connection, the associated question is: Can one formulate the energy eigenstate problems for ES modes in metal nano structures as well as for MS oscillations in ferrite particles based on the Schrödinger-like equation written, respectively, for an *ES-potential scalar wave function $\phi$* and a *MS-potential scalar wave function $\psi$*?

In this paper, we analyze the problem of energy quantization of ES resonances in subwavelength optical metallic structures with plasmon oscillations and MS resonances in subwavelength microwave ferrite particles with magnon oscillations. We start with a comparative study of these two types of quasistatic resonances. We show that with a formal assumption of existence of *quasistatic retardation effects*, caused by long range dipole-dipole correlations, one can clarify the notions of energy and currents (fluxes) for ES and MS waves and write the Schrödinger-like equations both for the ES and MS scalar wave functions. Initially stating that in subwavelength structures made of strongly dispersive materials the long range dipole-dipole correlation can be treated in terms of collective excitations of the system as a whole, we argue that between the electrostatic (with neglect of magnetic energy) and magnetostatic (with neglect of electric energy) eigenvalue problems there is a fundamental difference. We show that in a case of MS-potential scalar wave function $\psi$ one can observe *quasistatic retardation effects* and obtain a proper formulation of the energy eigenstate problem based on the Schrödinger-like equation. We show that in near-field regions of a MS-resonance ferrite particles there exist so-called magnetoelectric fields. Such quantized fields are distinguished by unique topological structures.

## II. QUASI-ELECTROSTATIC AND QUASI-MAGNETOSTATIC RESONANCES IN SUBWAVELENGTH STRUCTURES

### A. Quasi-electrostatic resonances in subwavelength optical metallic structures

For quasi-electrostatic resonances in subwavelength optical metallic structures, the electrostatic results are derived from an order expansion of Maxwell's equations. In this analysis, the problem is solved by expanding the dielectric constant of the metal nanoparticle $\varepsilon_m(\omega)$ at the resonance frequency $\omega_{res}$ in terms of the wave number in a host material $k_h = \sqrt{\varepsilon_h}\,\omega/c$ times the characteristic size of the nanoparticle $a$ as [1, 4, 5]

$$\varepsilon_m(\omega_{res}) = \varepsilon_m^{(0)} + (ak_h)\varepsilon_m^{(1)} + (ak_h)^2 \varepsilon_m^{(2)} + ... \tag{1}$$

With this expansion, one can identify the relative importance of the electromagnetic-field components in the electrostatic eigenvalue problem. The zeroth order term $\varepsilon_m^{(0)}$ describes the interaction between the electric fields and the dielectrics, which results in the electrostatic formulation ($c \to \infty$). In this case, the boundary value problems can be written in terms of ES-potential scalar function $\phi$. This function must be continuous and differentiable with respect to the normal to the boundary surface (Neumann-Dirichlet boundary conditions). The problem can be solved based on the electrostatic surface integral approach [4, 6, 7]. With formulation of proper boundary conditions for surface charges or surface dipoles over the metal-particle surface, for two eigenvalues $\left(\varepsilon_m^{(0)}\right)_i$ and $\left(\varepsilon_m^{(0)}\right)_k$ one obtains the orthogonality relations for electric-fields eigenfunctions $\vec{E}_i$ and $\vec{E}_k$. The electric permittivity of the metal and the surrounding (background) medium are related to the spectral eigenvalue $\gamma = \dfrac{\varepsilon_m(\omega) - \varepsilon_h}{\varepsilon_m(\omega) - \varepsilon_h}$, where the frequency $\omega$ is introduced as a parameter [4]. This technique is considered as being instrumental for development of an efficient numerical algorithm in an analysis of the coupling of plasmon resonance modes to the incident electromagnetic fields. Based on such a mathematical approach one can greatly simplify the eigenvalue problem.

Another way to solve the problem in the electrostatic formulation ($c \to \infty$) for plasmon resonances in a subwavelength structure is to solve the Poison's equation [8 – 11]:

$$\vec{\nabla}\cdot\left[\varepsilon(\vec{r},\omega)\vec{\nabla}\phi(\vec{r},\omega)\right] = 0. \tag{2}$$

Here the frequency $\omega$ is also introduced as a parameter. The mode expansion relies on the existence of an orthogonal and complete set of real electric-field eigenfunctions $\vec{E}_n$ and eigenvalues $\varepsilon_n$. For a composite consisting of the metal component with the permittivity $\varepsilon_m(\omega)$ and the ambient dielectric with the permittivity $\varepsilon_h$, the effective permittivity is calculated as

$$\varepsilon(\vec{r},\omega) = \varepsilon_h\left[1 - \theta(\vec{r})\left(1 - \frac{\varepsilon_m(\omega)}{\varepsilon_h}\right)\right], \tag{3}$$

where $\theta(\vec{r})$ is the characteristic function equal to 1 for $\vec{r}$ in the metal component and 0 for $\vec{r}$ in the dielectric. The eigenmodes $\phi_n(\vec{r})$ satisfy a generalized eigenproblem equation [9, 12, 13]:

$$\vec{\nabla}\cdot\left[\theta(\vec{r})\vec{\nabla}\phi_n(\vec{r})\right] = s_n \nabla^2 \phi_n(\vec{r}), \tag{4}$$

where $s_n(\omega) = \varepsilon_h/\left[\varepsilon_h - \varepsilon_m(\omega)\right]$. With use of Green function for Laplace equation $\nabla^2 G(\vec{r},\vec{r}') = -\delta^3(\vec{r}-\vec{r}')$, one can solve Eq. (4). Taking into account the Dirichlet-Neumann boundary conditions, we have the orthogonality relation

$$\langle \phi_n | \phi_m \rangle = \int_V \theta(\vec{r}) \vec{\nabla} \phi_n^* \cdot \vec{\nabla} \phi_m d^3 r . \tag{5}$$

The eigenmodes are normalized over a volume $V$ of a system: $\int_V |\vec{\nabla} \phi_n(\vec{r})|^2 d^3 r = 1$. Conventionally, the eigenmode normalization is $\int_V |\vec{E}_n(\vec{r})|^2 d^3 r = 1$. The spectral parameter can be represented also as $s_n = \int_V \theta(\vec{r}) |\vec{E}_n(\vec{r})|^2 d^3 r$.

It was shown [4] that in the expansion (1), the first-order correction to the to the dielectric constant at the resonance is zero, $\varepsilon_m^{(1)} = 0$. At the same time, with the second-order term in the expansion, $\varepsilon_m^{(2)}$, one corrects the electrostatic resonance modes by electromagnetic retardation. This term represents the electric induction created by the time-varying magnetic field, which interacts back on the surface charges. It means that a role of the magnetic field in plasmonic oscillations in metal nanoparticle becomes appreciable only when in an eigenvalue problem one deviates from the electrostatic approximation to the full-Maxwell-equation description.

The statements that first-order term of Eq. (1) is zero and that the magnetic field becomes appreciable only beyond the frames of the electrostatic approximation are questioned, however, in some works. It is inferred [1, 14, 15] that in an analysis of some effects one can use the first-order term $\varepsilon_m^{(1)}$ in electrostatic equations to predict magnetic fields. At a same time, it is assumed that this magnetic field, in itself, does not affect the electrostatic result. This statement is strongly misleading. First of all, with electrostatic equations one cannot derive the conductivity electric currents created by the oscillating charge distributions and thus use the Bio-Savart law for derivation of the magnetic field [14, 15]. The electrostatics and magnetostatics constitute two different parts of the electromagnetic theory and their interaction can be considered only in the frames of the full-Maxwell-equation description [3]. On the other hand, consideration of the magnetic field created by the electric displacement current (associated with the oscillating electric dipole moment) is also beyond a physical meaning in an analysis of electrostatic resonances. This becomes evident from the following analysis [16].

In a small (with sizes much less than a free-space electromagnetic wavelength) sample of a dielectric material with strong temporal dispersion (due to the plasmon resonance), one neglects a time variation of magnetic energy in comparison with a time variation of electric energy. In this case, the electromagnetic duality is broken and we have a system of three differential equations for the electric and magnetic fields

$$\nabla \cdot \vec{D} = 0, \tag{6}$$

$$\vec{\nabla} \times \vec{E} = 0, \tag{7}$$

$$\vec{\nabla} \times \vec{H} = \frac{\partial \vec{D}}{\partial t}, \tag{8}$$

In frames of the quasielectrostatic approximation, we introduce electrostatic-potential function $\phi(\vec{r}, t)$ excluding completely the magnetic displacement current: $\frac{\partial \vec{B}}{\partial t} = 0$. At the same time, from the Maxwell equation (the Ampere-Maxwell law), $\vec{\nabla} \times \vec{H} = \frac{\partial \vec{D}}{\partial t}$, we write that $\vec{\nabla} \times \frac{\partial \vec{H}}{\partial t} = \frac{\partial^2 \vec{D}}{\partial t^2}$. If a sample does not possess any magnetic anisotropy, we have $\frac{\partial^2 \vec{D}}{\partial t^2} = 0$. From this equation it follows that the electric field in small resonant objects vary linearly with time. This leads, however, to arbitrary large fields at early and late times, and is excluded on physical grounds. An evident conclusion suggests itself at once: the electric field in electrostatic resonances is a constant quantity. Such a conclusion contradicts the fact of temporally dispersive media and thus any resonant conditions. Thus, it becomes clear that the curl magnetic field appearing in plasmonic oscillations due to the Ampere-Maxwell law, not does not affect the electrostatic result. For this reason, such a magnetic field is a *non-observed* quantity.

The fact that for small metal particles the problem takes a mathematical form identical to that in electrostatics greatly simplifies the eigenvalue problem. However, this is only a mathematical approach. The quasielectrostatic model with neglecting any retardation effects, cannot accurately predict measured physical properties.

**B.    Quasi-magnetostatic resonances in subwavelength microwave ferrite structures**

Formally, one can see a certain resemblance between the electrostatic and magnetostatic resonances. For quasi-electrostatic resonances in subwavelength metal structures characterised by non-homogeneous scalar permittivity, Eq. (2) can be rewritten as

$$\nabla^2 \phi + \vec{\nabla} \cdot \left( \left( \varepsilon(\vec{r}) - 1 \right) \vec{\nabla} \phi \right) = 0. \tag{9}$$

At the same time, for quasi-magnetostatic resonances in subwavelength microwave ferrite structures with tensor permeability $\ddot{\mu}$, one has [2]:

$$\nabla^2 \psi + \vec{\nabla} \cdot \left( \left( \frac{\ddot{\mu}}{\mu_0} - \ddot{I} \right) \cdot \vec{\nabla} \psi \right) = 0. \tag{10}$$

Solutions of both these equations are harmonic functions. Nevertheless, it appears that in spite of a certain similarity between Eqs. (9) and (10), the physical properties of the oscillation spectra are fundamentally different in many aspects.

Alike to the ES resonances, MS resonances can be described by integral equation approaches. The integral equation approaches for MS resonances are applied with use of eigenvalue functions of magnetization [17, 18]. In a ferrite structure characterised by time-varying magnetization $\vec{m}(\vec{r},t)$, magnetostatic resonances are described by Poison's equation derived from the two equations $\vec{\nabla} \cdot \vec{B} = \vec{\nabla} \cdot (\vec{H} + \vec{m}) = 0$ and $\vec{H} = -\vec{\nabla} \psi$:

$$\nabla^2 \psi = -\rho_M, \tag{11}$$

where

$$\rho_M = -\vec{\nabla} \cdot \vec{m}. \tag{12}$$

The general solution for time-varying magnetic field is written as an integral transformation $\vec{H}(\vec{r},t) = \int_V \hat{\Gamma}(\vec{r},\vec{r}')\vec{m}(\vec{r}',t)d^3\vec{r}$ of the magnetization with the tensorial kernel operator $\hat{\Gamma}(\vec{r},\vec{r}')$, called a tensorial magnetostatic Green's function. The formal expression for the tensorial Green's function $\hat{\Gamma}(\vec{r},\vec{r}')$ is obtained by calculating spatial derivatives from the conventional ''Coulombs'' kernel [2]. The dynamical magnetization of the magnetic structure can be decomposed in series of spin wave eigenmodes $\vec{m}(\vec{r},t) = \sum_n a_n \vec{m}_n(\vec{r},t)e^{i\omega_n t}$, where $a_n$ are the amplitudes of the spin wave eigenmodes and on are their eigenfrequencies. The quantized frequencies $\omega_n$ of magnetostatic eigenmodes can be found from the solution of the integral equation $\lambda_n \vec{m}_n(\vec{r}) = \int_V \hat{\Gamma}(\vec{r},\vec{r}')\vec{m}_n(\vec{r}')d^3\vec{r}'$, where $\vec{m}_n(\vec{r})$ are the $n$-eigenmode spatial profile. The spin wave eigenfrequencies $\omega_n$ on of a magnetic element are the simple functions of the discrete eigenvalues $\lambda_n$ of the Fredholm integral with the symmetric magnetostatic kernel $\hat{\Gamma}(\vec{r},\vec{r}')$. The set of the derived eigenfunctions $\vec{m}_n(\vec{r})$ and eigenfrequencies $\omega_n$ is a complete solution of the problem of MS spin waves in a single magnetic element. The problem, however, does not have a definite analytical solution since the boundary conditions for variable magnetization at the edges of the pattern elements are not well defined. It is known that the usual electrodynamic boundary conditions leave the amplitude of dynamic magnetization at the boundaries of a sample undefined [2, 17, 18].

Together with integral equation approaches, one can formulate the spectral problem for MS resonances, also based on differential operators using MS-potential functions as eigenvalue functions. The spectral solutions are obtained from the second-order differential equation (10) – the Walker equation [2]. The MS wave function $\psi(\vec{r},t)$ is constructed in accordance with basic symmetry considerations for the sample geometry. There could be, for example, simple sine MS waves, Fourier–Bessel MS modes, and Fourier–Legendre MS modes. However, in this way of the spectral problem solution for MS resonances, application of proper boundary conditions appears as a very nontrivial question as well. It is known that in solving a boundary-value problem, which involves the eigenfunctions of a differential operator, the boundary conditions should be in definite correlation with the type of this differential operator [19, 20]. When we employ expansion in terms of orthogonal functions, we have to use the homogeneous Dirichlet-Neumann boundary conditions. For the fields expressed by MS-potential functions, this demand to have continuity of the functions together with their first derivatives on the sample boundaries (the Dirichlet-Neumann boundary conditions) can be different from the electrodynamic boundary conditions. The latter are expressed by the conditions of continuity of $\psi$ and a normal component of $\vec{B} = \hat{\mu} \cdot \vec{H} = -\hat{\mu} \cdot \vec{\nabla}\psi$. For

$$\hat{\mu} = \mu_0 \begin{bmatrix} \mu & i\mu_a & 0 \\ -i\mu_a & \mu & 0 \\ 0 & 0 & 1 \end{bmatrix}, \qquad (13)$$

the Walker equation in Cartesian coordinates is

$$\mu\left(\frac{\partial^2\psi}{\partial^2 x}+\frac{\partial^2\psi}{\partial^2 y}\right)+\frac{\partial^2\psi}{\partial^2 z}=0. \qquad (14)$$

Evidently, the Walker equation does not contain the off-diagonal component of the permeability tensor. At the same time, the off-diagonal component $\mu_a$ can appear in a boundary conditions for a normal component of $\vec{B}$.

The boundary conditions for dynamic magnetization and it derivatives are not the electrodynamic boundary conditions. On the other hand, because of magnetic gyrotropy, electrodynamic boundary conditions for the $\psi$-function solutions are not Dirichlet-Neumann boundary conditions. These aspects may question the validity of the boundary-value-problem solutions written in terms of MS-potential scalar function. The question on quasistatic magnonic eigenproblem with complete-set eigenstates becomes essentially crucial when we are talking about combination of a microwave cavity and a subwavelength YIG ferromagnet specimen. Such a microwave structure represents a promising path towards the ultrastrong-coupling regime of QED. MS oscillations in YIG spheres were observed and analysed long ago [21, 22]. While for a ferrite sphere one sees a few broad absorption peaks, the MS resonances in a quasi-2D ferrite disk are presented with the spectra of multiresonance sharp peaks. There are very rich spectra of both, Fano and Lortenzian, types of the peaks. A thin-film ferrite disk with MS oscillations embedded in a microwave waveguide or microwave cavity appears as an open hi-Q resonator [23 – 25]. In our further analysis of MS resonances, we will consider mainly the quasi-2D ferrite disk samples.

### III. ENERGY AND CURRENTS (FLUXES) OF ES AND MS MODES

The effects of quantum coherence involving a macroscopic degree of freedom, and occurring in systems far larger than individual atoms are one of the topical fields in modern physics [26]. Because of material dispersion, a phenomenological approach to macroscopic quantum electrodynamics, where no canonical formulation is attempted, is used. There is an evidence that macroscopic systems can under appropriate conditions, be in quantum states, which are linear superpositions of states with different macroscopic properties. For subwavelength structures made of strongly dispersive materials long range dipole-dipole correlation can be treated in terms of collective excitations of the system as a whole. At such an occasion, potential functions, used in the quasistatic-resonance spectral problem analysis, can at some cases, be introduced as scalar *wave* functions.

Recent years we have witnessed of development of a new field of optics – quantum plasmonics – which combines the advantages of plasmonics and quantum electronics [27]. Considering a metamaterial structure composed by plasmon-resonance metal nanoparticles embedded in a dielectric host material, Bergman and Stockman introduced a notion of a quantum generator for surface *plasmon quanta* – the spaser [9, 12, 13]. In the spaser, nanoscale resonators are employed for controlling stimulated emission regimes, where the near-field feedback replaces traveling phase behavior in a photon cavity. In a quasistatic approximation used for the spaser analysis, it is assumed that the surface-plasmon (SP) eigen modes are non-propagating modes. This approximation means that there is zero current carried by any eigenmode. Quantization of the SP system is obtained based on the Hamiltonian used for a temporally dispersive dielectric medium $H=\frac{1}{4\pi T}\int \frac{d\left[\omega\varepsilon(\vec{r},\omega)\right]}{d\omega}\vec{E}(\vec{r},\omega)\vec{E}(\vec{r},-\omega)\frac{d\omega}{2\pi}d^3r$ [12, 28]. In the quasielectrostatic regime [for $\vec{E}(\vec{r},t)=-\vec{\nabla}\phi(\vec{r},t)$], the electric field operator is expanded in a series of the eigenstates $\phi_n(\vec{r})$:

$$\hat{\vec{E}} = -\sum_n A_n \vec{\nabla} \phi_n(\vec{r})(\hat{a}_n^\dagger + \hat{a}_n),\tag{15}$$

where $A_n$ is the mode amplitude and $\hat{a}_n^\dagger$ and $\hat{a}_n$ are, respectively, the creation and annihilation operators of the state $\phi_n(\vec{r})$. The validity of the expansion is relayed on an assertion that the eigenmodes $\phi_n(\vec{r})$ satisfy a generalized eigenproblem equation (4). It is stated that with this expansion, the quantized Hamiltonian takes the standard harmonic oscillator form [12, 13, 29].

It is known that the quantized form of the SP oscillations are bosons and the structure of the *total electromagnetic energy* of surface plasmon polaritons have a harmonic oscillator form. The fields of these SPs are quantized by the association of a quantum mechanical oscillator for each *mode wave number* [27]. The quantum harmonic oscillator is the quantum mechanical analog of the classical harmonic oscillator. For a sinusoidal driving force, an analogy of such an oscillator is RLC circuits (resistor-inductor-capacitor). However, the questions arise: Whether the SPs can be considered as the bosons in the *electrostatic* representation? Does the quantized *electrostatic* Hamiltonian take the standard harmonic-oscillator form for eigenmodes with zero mode wave numbers? The SP resonances are driven by external electric fields. In a quasistatic description, used in the Bergman-Stockman model, zero currents carried by eigenmodes are assumed. Thus, to an approximate extent, we have a system which is similar to a resistor–capacitor circuit (RC circuit) driven by an external voltage. Definitely, this is not a model of a classical harmonic oscillator.

The spaser concept paves the way for the creation of strong coherent plasmonic fields at the nanoscale. A significant number of generating spasers have been reported in recent experiments [30 – 32]. Nevertheless, the question arises whether the Bergman-Stockman *quasistatic* theory is sufficient for explanation of the observed spaser effects. In spasers, nanoscale resonators are employed for controlling stimulated emission regimes, where the electromagnetically near-field feedback replaces traveling phase behavior in a photon cavity. Is it correct that in the theory, such a near-field feedback lacks any retardation behavior? In literature, we can see a certain criticism of the quasistatic model of spasers with pointing out the necessity to extend the electrostatic limit by the electromagnetic radiation correction. It is discussed that quasistatic approach is inherently unsuitable due to consideration only one decay channel for the resonance through resistive losses in the metal [33, 34].

How can one include retardation effects still staying in frames of the quasistatic description of oscillations in a subwavelength particle? The fact that the electromagnetic retardation effects in the plasmonic or magnonic oscillations appear only when the particle sizes are comparable with the free-space electromagnetic wavelength raises the question on the possibility of existence of non-Maxwellian propagation-wave behaviors for the quasistatic-resonance processes. Such wave processes, if any, should be considered as currents (fluxes) expressed by scalar wave functions $\phi(\vec{r},t)$ for ES waves and scalar wave functions $\psi(\vec{r},t)$ for MS waves. A formal analysis given below clarifies this question. It is shown that in this case, one has a possibility to formulate the energy eigenstate boundary problem with scalar-wave eigenfunctions based on the Schrödinger-like equation.

In a functional analysis, a spectral theorem gives conditions for diagonalization of a linear operator. Suppose that for steady states of the system we have a time-independent equation

$$\hat{\mathcal{F}}\vartheta = \mathcal{E}\vartheta,\tag{16}$$

where $\hat{\mathcal{F}}$ is a Hermitian second-order differential operator, $\vartheta$ is a certain scalar wave function and $\mathcal{E}$ is a real quantity. For a domain of volume *V* restricted by surface *S*, a double integration by parts gives

$$\int_V \left(\hat{\mathcal{F}}\vartheta\right)\vartheta^* dV = \int_V \left(\hat{\mathcal{F}}\vartheta^*\right)\vartheta dV + \int_S \mathcal{P}\left(\vartheta,\vartheta^*\right) dS . \tag{17}$$

For homogeneous boundary conditions, one has $\int_S \mathcal{P}\left(\vartheta,\vartheta^*\right) dS = 0$ and operator $\hat{\mathcal{F}}$ becomes a self-adjoint operator. If $\hat{\mathcal{F}}$ is a Laplace operator ($\hat{\mathcal{F}} = -K\nabla^2$, where *K* is a coefficient), the homogeneous boundary conditions look as $\int_S \left(\vartheta\vec{\nabla}\vartheta^* - \vartheta^*\vec{\nabla}\vartheta\right) dS = 0$. In this case, for two eigenvalues $\mathcal{E}_m, \mathcal{E}_n$ and, correspondingly, for two eigenfunctions $\vartheta_m, \vartheta_n$ one has

$$\left(\mathcal{E}_m - \mathcal{E}_n\right)\vartheta_m \vartheta_n^* = K\left(\vartheta_m \nabla^2 \vartheta_n^* - \vartheta_n^* \nabla^2 \vartheta_m\right) = K\vec{\nabla}\cdot\left(\vartheta_m \vec{\nabla}\vartheta_n^* - \vartheta_n^* \vec{\nabla}\vartheta_m\right). \tag{18}$$

Integration over the entire volume *V* gives the orthonormality conditions:

$$\left(\mathcal{E}_m - \mathcal{E}_n\right)\int_V \vartheta_m \vartheta_n^* dV = 0 . \tag{19}$$

In quantum mechanics, $\hat{\mathcal{F}}$ is the operator associated with energy – the Hamiltonian – and $\mathcal{E}$ is an observable quantity of energy. The operation of the Hamiltonian on the scalar wave function is the time-independent Schrödinger equation. The wavefunction of an electron can be decomposed with a complete set of scalar eigenfunctions, which obey the time-independent Schrödinger equation. The term $\vec{\mathcal{J}} \propto \vartheta\vec{\nabla}\vartheta^* - \vartheta^*\vec{\nabla}\vartheta$ is associated with the quantum motion of the charges. The form of the electron wave function determines the current $\vec{\mathcal{J}}$, which is called the flow probability. In stationary states (the energy eigenstates), the probability current is spatially uniform or zero. This means that in a stationary state, $\vec{\nabla}\cdot\vec{\mathcal{J}} = 0$ [35]. The above the well-known procedure can be formally used for other types of scalar wave functions – the scalar wave functions of the ES and MS resonant structures. As we will show, in a case of both ES and MS resonances one can introduce a certain notion of current $\vec{\mathcal{J}}$, which is determined by the term like $\vartheta\vec{\nabla}\vartheta^* - \vartheta^*\vec{\nabla}\vartheta$ and has a meaning of the power flow density for the ES and MS waves.

Let us assume that in a temporally dispersive dielectric medium we have waves, which are determined by a scalar wave function $\phi(\vec{r},t)$. These waves are pure electrostatic, without any accumulation of magnetic energy. A retardation process is due to dipole-dipole interaction of electric dipoles. Similarly, we assume that in a temporally dispersive magnetic medium the wave process, being pure magnetostatic, is determined by a scalar wave function $\psi(\vec{r},t)$ without any accumulation of electric energy. Here have a retardation process is due to dipole-dipole interaction of magnetic dipoles. Suppose that in both cases, the waves propagate in anisotropic media and the media do not contain any free charges and conductivity currents.

Admitting that the media has small losses, we have the following continuity equation for ES waves propagating at frequency $\omega$

$$-\frac{i\omega}{4}\vec{\nabla}\cdot\left(\phi\vec{D}^* - \phi^*\vec{D}\right) = -\frac{i\omega}{4}\left[\vec{D}\cdot\left(\left(\vec{\varepsilon}^*(\omega)\right)^{-1}\cdot\vec{D}^*\right) - \vec{D}^*\cdot\left(\left(\vec{\varepsilon}(\omega)\right)^{-1}\cdot\vec{D}\right)\right] = \left\langle\frac{\partial w_{el}}{\partial t}\right\rangle_{abs}. \quad (20)$$

On the right-hand side of this equation we have the average density of electric losses taken with an opposite sign. In derivation of Eq. (20) we used the following relations: $\vec{E} = -\vec{\nabla}\phi$, $\vec{D} = \vec{\varepsilon}\cdot\vec{E} = -\vec{\varepsilon}\cdot\vec{\nabla}\phi$, and $\vec{\nabla}\cdot\vec{D} = 0$. The average density of electric losses is defined as

$$\left\langle\frac{\partial w_{el}}{\partial t}\right\rangle_{abs} = \frac{1}{2}i\omega\vec{E}^*\cdot\left(\vec{\varepsilon}^{ah}(\omega)\cdot\vec{E}\right), \quad (21)$$

where in the permittivity tensor $\vec{\varepsilon}$ superscript *ah* means 'anti-Hermitian'. The energy balance equation for monochromatic MS waves in a lossy ferrite medium we write as

$$-\frac{i\omega}{4}\vec{\nabla}\cdot\left(\varphi\vec{B}^* - \varphi^*\vec{B}\right) = -\frac{i\omega}{4}\left[\vec{B}\cdot\left(\left(\vec{\mu}^*(\omega)\right)^{-1}\cdot\vec{B}^*\right) - \vec{B}^*\cdot\left(\left(\vec{\mu}(\omega)\right)^{-1}\cdot\vec{B}\right)\right] = \left\langle\frac{\partial w_{mag}}{\partial t}\right\rangle_{abs}. \quad (22)$$

In derivation of Eq. (22) we used the following relations: $\vec{H} = -\vec{\nabla}\psi$, $\vec{B} = \vec{\mu}\cdot\vec{H} = -\vec{\mu}\cdot\vec{\nabla}\psi$, and $\vec{\nabla}\cdot\vec{B} = 0$. The average density of magnetic losses is defined as

$$\left\langle\frac{\partial w_{mag}}{\partial t}\right\rangle_{abs} = \frac{1}{2}i\omega\vec{H}^*\cdot\left(\vec{\mu}^{ah}(\omega)\cdot\vec{H}\right), \quad (23)$$

where in the permeability tensor $\vec{\mu}$ superscript *ah* means 'anti-Hermitian'.

With the above analysis, we can see that if in a anisotropic dielectric medium there exist the ES waves described by a scalar wave function $\phi(\vec{r},t)$, the power flow density is viewed as a certain current density:

$$\vec{\mathcal{J}} = \frac{i\omega}{4}\left(\phi\vec{D}^* - \phi^*\vec{D}\right). \quad (24)$$

This power flow can be observed because of dipole-dipole interaction of electric dipoles. Similarly, for MS waves in a ferrite medium, described by a scalar wave function $\psi(\vec{r},t)$, we have the power flow density which can be viewed as a current density:

$$\vec{\mathcal{J}} = \frac{i\omega}{4}\left(\psi\vec{B}^* - \psi^*\vec{B}\right). \quad (25)$$

Such a power flow can appear because of dipole-dipole interaction of magnetic dipoles.

Let, in a particular case, both the dielectric and magnetic media be isotropic with the constitutive relations $\vec{D} = \varepsilon\vec{E}$ and $\vec{B} = \mu\vec{H}$. For ES and MS waves, we have, respectively,

$$\vec{\mathcal{J}} = -\frac{i\omega\varepsilon}{4}\left(\phi\vec{\nabla}\phi^* - \phi^*\vec{\nabla}\phi\right) \quad (26)$$

and

$$\vec{\mathcal{J}} = -\frac{i\omega\mu}{4}\left(\psi\vec{\nabla}\psi^* - \psi^*\vec{\nabla}\psi\right). \quad (27)$$

Assume that using the Laplace operator and Dirichlet-Neumann boundary conditions in ES and MS spectral problems, we can decompose the wave functions $\phi(\vec{r},t)$ and $\psi(\vec{r},t)$ with a complete set of scalar eigenfunctions, obeying the time-independent Schrödinger-like equation. It this case, we are able to rewrite the orthonormality relation (19) as

$$\left(\mathcal{E}_m - \mathcal{E}_n\right)\int_V \phi_m \phi_n^* dV = 0 \quad (28)$$

for ES resonances and

$$\left(\mathcal{E}_m - \mathcal{E}_n\right)\int_V \psi_m \psi_n^* dV = 0 \quad (29)$$

for MS resonances.

## IV. MS MAGNONS IN FERRITE-DISK RESONATORS

Eqs. (24 – 29) were obtained in an assumption that in a subwavelength specimen we have propagation of quasistatic bulk waves. This not the case of SP resonances in subwavelength optical metallic structures. In such structures, no retardation processes characterized by the electric dipole-dipole interaction and described exclusively by electrostatic wave function $\phi(\vec{r},t)$ take place. There is no possibility to describe these resonances by the Schrödinger-equation energy eigenstate problem. Nevertheless, for MS resonances in ferrite specimens we have bulk wave process, which are determined by a scalar wave function $\psi(\vec{r},t)$ [2, 28]. Due to retardation processes caused by the magnetic dipole-dipole interaction in a subwavelength ferrite particle, we have a possibility to formulate the energy eigenstate boundary problem with scalar-wave eigenfunctions $\psi(\vec{r},t)$ based on the Schrödinger-like equation. Such a behavior can be obtained in a ferrite particle in a form of a quasi-2D disk. As we pointed out above, a ferrite disk is distinguished from a ferrite sphere by multiresonance sharp-peak spectra of MS oscillations. The oscillations in a quasi-2D ferrite disk, analyzed as spectral solutions for the MS-potential scalar wave function $\psi(\vec{r},t)$, has evident quantum-like attributes. Quantized forms of such matter oscillations we call the MS magnons or the magnetic-dipolar-mode (MDM) magnons. The macroscopic nature of MDMs, involving the collective motion of a many-body system of precessing electrons, does not destroy a quantum behavior. The long-range dipole-dipole correlation in positions of electron spins can be treated in terms of collective excitations of a system as a whole.

### A. Energy eigenstates of MS magnons

We analyse functions $\psi$ within the space of square integrable functions. For an open quasi-2D ferrite disk normally magnetized along the *z* axis, we can use separation of variables in the spectral problem solution [36 – 38]. In cylindrical coordinate system $(z, r, \theta)$, the solution is represented as

$$\psi_{p,v,q} = A_{p,v,q}\xi_{p,v,q}(z)\tilde{\eta}_{v,q}(r,\theta),\tag{30}$$

where $A_{p,v,q}$ is a dimensional amplitude coefficient, $\xi_{p,v,q}(z)$ is a dimensionless function of the MS-potential distribution along $z$ axis, and $\tilde{\eta}_{v,q}(r,\theta)$ is a dimensionless membrane function. The membrane function $\tilde{\eta}$ is defined by a Bessel-function order $v$ and a number of zeros of the Bessel function corresponding to a radial variations $q$. The dimensionless "thickness-mode" function $\xi(z)$ is determined by the axial-variation number $p$ [36 – 38].

Suppose that a mode number $n$ holds a certain set of quantum numbers $v, q, p$. The solution for the mode $n$ is represented as

$$\psi_n = A_n \xi_n(z)\tilde{\eta}_n(r,\theta),\tag{31}$$

In a further analysis we will use the subscript $\perp$ to denote differentiation over the in-plane $r,\theta$ coordinates and subscript $\parallel$ to denote differentiation along the disk axis $z$. Also, we will apply subscript $\perp$ for the vector and tensor components laying in the $r,\theta$ plane. Using the separation of variables we write the Walker equation (10) in the following form:

$$\mu\nabla_{\perp}^2\psi + \nabla_{\parallel}^2\psi = 0,\tag{32}$$

where $\mu$ is a diagonal component of the permeability tensor (13). An axial variation of a function $\xi(z)$ is defined by a solution of a boundary value problem with the function description inside a ferrite as $\xi_n(z) \propto \cos\beta_n z + \dfrac{1}{\sqrt{-\mu_n}}\sin\beta_n z$, outside a ferrite as $\xi_n(z) \propto e^{\pm\alpha_n z}$ (where $\alpha_n = \dfrac{1}{\sqrt{-\mu_n}}\beta_n$), and with using the homogenious boundary conditions on the disk surface planes $z = 0$ and $z = d$:

$$(\psi_n)_{\substack{z=0^-\\z=d^-}} - (\psi_n)_{\substack{z=0^+\\z=d^+}} = 0,$$

$$\left[\left(\frac{\partial\xi_n}{\partial z}\right)_n\right]_{\substack{z=0^-\\z=d^-}} - \left[\left(\frac{\partial\xi_n}{\partial z}\right)_n\right]_{\substack{z=0^+\\z=0^+}} = 0.\tag{33}$$

For the disk geometry, the energy eigenvalue problem for MS modes is defined by the differential equation [36 – 39]

$$\hat{G}_{\perp}\tilde{\eta}_n = E_n\tilde{\eta}_n,\tag{34}$$

where $\hat{G}_{\perp}$ is a two-dimensional differential operator and $E_n$ is interpreted as density of accumulated magnetic energy of mode $n$. The operator $\hat{G}_{\perp}$ and quantity $E_n$ are defined as

$$\hat{G}_{\perp} = \frac{g_n \mu_0}{4} \mu_n \nabla_{\perp}^2, \tag{35}$$

$$E_n = \frac{g_n \mu_0}{4} \left(\beta_{z_n}\right)^2. \tag{36}$$

Here $g_n$ is a dimensional normalization coefficient for mode $n$ and $\beta_{z_n}$ is the propagation constant of mode $n$ along the disk axis $z$. The parameter $\mu_n$ is to be regarded as an eigenvalue. Outside a ferrite $\mu_n = 1$. The operator $\hat{G}_{\perp}$ is a self-adjoint operator only for negative quantities $\mu_n$ in a ferrite.

For self-adjointness of operator $\hat{G}_{\perp}$, the membrane function $\tilde{\eta}_n(r,\theta)$ must be continuous and differentiable with respect to the normal to lateral surface of a ferrite disk. The homogeneous boundary conditions – the Neumann-Dirichlet boundary conditions – for the membrane function should be [36 – 41]:

$$\left(\tilde{\eta}_n\right)_{r=\mathcal{R}^-} - \left(\tilde{\eta}_n\right)_{r=\mathcal{R}^+} = 0 \tag{37}$$

and

$$\mu \left(\frac{\partial \tilde{\eta}_n}{\partial r}\right)_{r=\mathcal{R}^-} - \left(\frac{\partial \tilde{\eta}_n}{\partial r}\right)_{r=\mathcal{R}^+} = 0, \tag{38}$$

where $\mathcal{R}$ is a disk radius.

MDM oscillations in a ferrite disk are described by real eigenfunctions: $\left(\tilde{\eta}_{-\beta}\right)_n = \left(\tilde{\eta}_{\beta}^*\right)_n$. For modes $n$ and $n'$, the orthogonality conditions are expressed as

$$\int_{S_c} \left(\tilde{\eta}_{\beta}\right)_n \left(\tilde{\eta}_{-\beta}\right)_{n'} dS = \int_{S_c} \tilde{\eta}_n \tilde{\eta}_{n'}^* dS = \delta_{nn'}. \tag{39}$$

where $S_c$ is a circular cross section of a ferrite-disk region and $\delta_{nn'}$ is the Kronecker delta.

The spectral problem gives the energy orthogonality relation for MDMs:

$$\left(E_n - E_{n'}\right) \int_{S_c} \tilde{\eta}_n \tilde{\eta}_{n'}^* dS = 0. \tag{40}$$

Since the space of square integrable functions is a Hilbert space with a well-defined scalar product, we can introduce a basis set. A dimensional amplitude coefficient in Eq. (30) we write as $A_n = c' a_n$, where $c'$ is a dimensional unit coefficient and $a_n$ is a normalized dimensionless amplitude. The normalized scalar-wave membrane function $\tilde{\eta}$ can be represented as

$$\tilde{\eta} = \sum_n a_n \tilde{\eta}_n. \tag{41}$$

The amplitude is defined as

$$|a_n|^2 = \left| \int_{S_c} \tilde{\eta} \tilde{\eta}_n^* dS \right|^2, \qquad (42)$$

The mode amplitude in Eq. (40) can be interpreted as the probability to find a system in a certain state *n*. Normalization of membrane function is expressed as

$$\sum_n |a_n|^2 = 1. \qquad (43)$$

The above analysis of discrete-energy eigenstates of the MS-wave oscillations, resulting from structural confinement in a normally magnetized ferrite disk, was based on a continuum model. Using the principle of wave–particle duality, one can describe this oscillating system as a collective motion of quasiparticles. These quasiparticles are called "light" magnons (lm). There are the MS magnons and the meaning of the term "light" arises from the fact that the effective masses of these quasiparticles are much less, than the effective masses of "real" ("heavy") magnons – the quasiparticles existing due to the exchange interaction. The spatial scale of the exchange interaction is much less than the MS-wave wavelength in our structures and the "magnetic stiffness" is characterized by the "weak" dipole-dipole interaction [2, 37].

An expression for the effective mass of the "light" magnon we derive from the following consideration. The MS-potential wavefunction $\psi$ entirely defines the MS-oscillation states of a ferrite disk. Thereby, representation of this function in a certain time moment not only describes the system behavior at the present moment, but defines the behaviour in all future time moments. It means that at every time moment, a time derivative $\frac{\partial \psi}{\partial t}$ should be defined by itself function $\psi$ at the same time moment. Moreover, because of the principle of superposition this relation has to be linear. In a general form, we can write that [35]

$$\frac{\partial \psi}{\partial t} = i\hat{Q}\psi, \qquad (44)$$

where $\hat{Q}$ is a certain linear operator. From this equation it can be shown that for orthonormalized basic vectors operator $\hat{Q}$ is a self-conjugate differential operator. Thus, equation (44) is a wave equation for complex scalar wavefunction $\psi$ [35].

Let us represent the function $\psi$ as a quasi-monochromatic wave propagating along a disk axis:

$$\psi(z,t) = \psi^{max}(z,t) e^{i(\omega t - \beta z)}, \qquad (45)$$

where a complex amplitude $\psi^{max}(z,t)$ is a smooth function of the longitudinal coordinate and time, so that

$$\left| \beta^{-1} \frac{\partial \psi^{max}}{\partial z} \right| \ll \psi^{max} \qquad (46)$$

and

$$\left| \omega^{-1} \frac{\partial \psi^{max}}{\partial t} \right| \ll \psi^{max} . \qquad (47)$$

The situation of the quasi-monochromatic behavior can be realized, in particular, by means of a time-dependent bias magnetic field slowly varying with respect to the Larmor frequency. In this case, a spin-polarized ensemble will adiabatically follow the bias magnetic field and the resulting energy of interaction with a bias field becomes time dependent.

For a quasi-monochromatic wave, Eq. (44) can be written as

$$\frac{\partial \psi}{\partial t} = i \frac{\omega}{\beta^2} \nabla_\parallel^2 \psi . \qquad (48)$$

The form of operator $\hat{Q} \equiv i \frac{\omega}{\beta^2} \nabla_\parallel^2 \psi$ follows from the 'stationary-state' conditions. When we compare Eq. (48) with the Schrödinger equation for "free particles" we get the expression for the effective mass of the "light" magnon for a monochromatic MS mode:

$$\left[ m_{eff}^{(lm)} \right]_n = \frac{\hbar}{2} \frac{\beta_n^2}{\omega} . \qquad (49)$$

Expression (49) looks very similar to the effective mass of the "heavy" magnon for spin waves with a quadratic character of dispersion [2]. The MS "light" magnons can be considered as a free particle with a mass: $\left[ m_{eff}^{(lm)} \right]_n$. Hamiltonian of sich a free particle is $\hat{H} = -\frac{\hbar^2}{2 \left[ m_{eff}^{(lm)} \right]_n} \nabla_\parallel^2$. For homogeneous pressesion of magnetization the energy is $E = \hbar \omega_0$, where, for a thin cylinder, $\omega_o = \gamma \left( H_o^{external} + 2\pi M_0 \right)$. A surplus of energy $\Delta E = \hbar \left( \omega_n - \omega_0 \right)$ can be interpreted as a magnon potential energy in an external and demagnetization magnetic fields. In a quasi-2D disk, the MS "light" magnons are "flat-mode" quasiparticles at a reflexively-translational motion behavior between the lower ($z = 0$) and upper ($z = d$) planes of a ferrite disk.

It is important to note that while in quantum mechanics problems, Schrödinger equation is written for 1D structure, in our case we have a 2D problem. In the spectral problem solution for membrane-function "flat modes" we use parameters obtained from the spectral problem solution for "thickness modes". However, in a case of a ferrite disk with a very small thickness/diameter ratio, the spectrum of "thickness modes" is very rare compared to the dense spectrum of "flat modes". The entire spectrum of "flat modes" is completely included in the wavenumber region of a fundamental "thickness mode". This means that the spectral properties of a resonator can be entirely described based on consideration of only a fundamental "thickness mode" and the problem appears as a quasi one dimensional [38].

**B. Power flow densities**

With use of separation of variables and taking into account a form of tensor $\vec{\mu}$ [see Eq. (13)], we decompose a magnetic flux density ($\vec{B} = \vec{\mu} \cdot \vec{H} = -\vec{\mu} \cdot \vec{\nabla} \psi$) by two components:

$$\vec{B} = \vec{B}_\perp + \vec{B}_\parallel. \tag{50}$$

The component $\vec{B}_\perp$ are given as

$$\vec{B}_\perp = -\vec{\mu}_\perp \cdot \vec{\nabla}_\perp \psi = -A_n \xi_n(z) \left[ \vec{\mu}_\perp \cdot \vec{\nabla}_\perp \tilde{\eta}_n(r,\theta) \right] \vec{e}_\perp, \tag{51}$$

where $\vec{e}_\perp$ is a unit vector laying in the $r, \theta$ plane, and

$$\vec{\mu}_\perp = \mu_0 \begin{bmatrix} \mu & i\mu_a \\ -i\mu_a & \mu \end{bmatrix}. \tag{52}$$

For the component $\vec{B}_\parallel$ we have

$$\vec{B}_\parallel = -\mu_0 \cdot \vec{\nabla}_\parallel \psi = -\mu_0 A_n \frac{\partial \xi_n(z)}{\partial z} \tilde{\eta}_n(r,\theta) \vec{e}_z, \tag{53}$$

where $\vec{e}_z$ is a unit vector directed along the $z$ axis.

The above representations allow considering, respectively, two components of the power flow density (current density). For mode $n$, we can write Eq. (25) as

$$\vec{\mathcal{J}}_n = \left(\vec{\mathcal{J}}_\perp\right)_n + \left(\vec{\mathcal{J}}_\parallel\right)_n, \tag{54}$$

where $\left(\vec{\mathcal{J}}_\perp\right)_n = \frac{i\omega}{4} \left[ \psi_n \left(\vec{B}_\perp^*\right)_n - \psi_n^* \left(\vec{B}_\perp\right)_n \right]$ and $\left(\vec{\mathcal{J}}_\parallel\right)_n = \frac{i\omega}{4} \left[ \psi_n \left(\vec{B}_\parallel^*\right)_n - \psi_n^* \left(\vec{B}_\parallel\right)_n \right]$. Along every of the coordinates $\vec{r}, \vec{\theta}$, and $\vec{z}$, we have power flows (currents):

$$\begin{aligned} \left(\vec{\mathcal{J}}_r\right)_n &= -\frac{i\omega}{4} |A_n|^2 |\xi_n|^2 \mu_0 \left\{ \tilde{\eta}_n \left( \mu \frac{\partial \tilde{\eta}_n}{\partial r} + i\mu_a \frac{1}{r} \frac{\partial \tilde{\eta}_n}{\partial \theta} \right)^* - \tilde{\eta}_n^* \left( \mu \frac{\partial \tilde{\eta}_n}{\partial r} + i\mu_a \frac{1}{r} \frac{\partial \tilde{\eta}_n}{\partial \theta} \right) \right\} \\ &= -\frac{i\omega}{4} |A_n|^2 |\xi_n|^2 \mu_0 \left\{ \mu \left[ \tilde{\eta}_n \left( \frac{\partial \tilde{\eta}_n}{\partial r} \right)^* - \tilde{\eta}_n^* \frac{\partial \tilde{\eta}_n}{\partial r} \right] + i\mu_a \frac{1}{r} \left[ \tilde{\eta}_n \left( \frac{\partial \tilde{\eta}_n}{\partial \theta} \right)^* - \tilde{\eta}_n^* \frac{\partial \tilde{\eta}_n}{\partial \theta} \right] \right\} \vec{e}_r, \end{aligned} \tag{55}$$

$$\begin{aligned} \left(\vec{\mathcal{J}}_\theta\right)_n &= -\frac{i\omega}{4} |A_n|^2 |\xi_n|^2 \mu_0 \left\{ \tilde{\eta}_n \left( -i\mu_a \frac{\partial \tilde{\eta}_n}{\partial r} + \mu \frac{1}{r} \frac{\partial \tilde{\eta}_n}{\partial \theta} \right)^* - \tilde{\eta}_n^* \left( -i\mu_a \frac{\partial \tilde{\eta}_n}{\partial r} + \mu \frac{1}{r} \frac{\partial \tilde{\eta}_n}{\partial \theta} \right) \right\} \\ &= -\frac{i\omega}{4} |A_n|^2 |\xi_n|^2 \mu_0 \left\{ \mu \frac{1}{r} \left[ \tilde{\eta}_n \left( \frac{\partial \tilde{\eta}_n}{\partial \theta} \right)^* - \tilde{\eta}_n^* \frac{\partial \tilde{\eta}_n}{\partial \theta} \right] - i\mu_a \left[ \tilde{\eta}_n \left( \frac{\partial \tilde{\eta}_n}{\partial r} \right)^* - \tilde{\eta}_n^* \frac{\partial \tilde{\eta}_n}{\partial r} \right] \right\} \vec{e}_\theta, \end{aligned} \tag{56}$$

$$\left(\vec{\mathcal{J}}_z\right)_n = -\frac{i\omega}{4}\mu_0|A_n|^2|\tilde{\eta}|^2\left[\xi_n\left(\frac{\partial\xi_n}{\partial z}\right)^* - \xi_n^*\frac{\partial\xi_n}{\partial z}\right]\vec{e}_z, \tag{57}$$

where $\vec{e}_r, \vec{e}_\theta$, and $\vec{e}_z$ are the unit vectors.

We can see that in in-plane coordinates, $r, \theta$, we have both real and imaginary power flows. That is the radial and azimuth power flows are complex quantities: $\mathcal{J}_r = (\mathcal{J}_r)_{real} + i(\mathcal{J}_r)_{imag}$ and $\mathcal{J}_\theta = (\mathcal{J}_\theta)_{real} + i(\mathcal{J}_\theta)_{imag}$. The imaginary parts of the flows shown in Eqs. (55), (56) are due to the material gyrotropy. In an electromagnetic theory, a reactive energy is related to the non-propagating (evanescent) fields. For the wave propagating along a certain direction with a real power flow, the imaginary part of the Poynting vector indicates a reactive power flux that oscillates back and forth in the transverse direction [3].

Let a disk plane region $S = \pi r^2$ be bounded by a circle $L = 2\pi r$, where $0 < r \leq \mathcal{R}$. For a two-dimensional vector field $\vec{\mathcal{J}}_\perp(\mathcal{J}_r, \mathcal{J}_\theta)$, the circulation-form Green's theorem is written in polar coordinates as

$$\iint_S \left(\frac{1}{r}\frac{\partial}{\partial r}(r\mathcal{J}_\theta) - \frac{1}{r}\frac{\partial \mathcal{J}_r}{\partial \theta}\right)ds = \oint_L (\mathcal{J}_r dr + r\mathcal{J}_\theta d\theta) \tag{58}$$

where $dS = rdrd\theta$. In a case of real power flows, $\tilde{\eta}$ is a singlevalued function. Such a membrane function is an azimuthal standing wave: it has an integer number of wavelengths around the circle. So, a circulation of gradient $\vec{\nabla}_\theta \tilde{\eta}$ along contour $L$ is equal to zero. In this instance, we have $\tilde{\eta}^* = -\tilde{\eta}$. Using Eqs. (55) and (56), we can easily show that for real power flows, an integrand in left-hand side of Eq. (58) is equal to zero. If such a double integrand of a vector field $\vec{\mathcal{J}}_\perp(\mathcal{J}_r, \mathcal{J}_\theta)$ is zero then this field is said to be irrotational. Since we have $\oint_L (\mathcal{J}_r dr + r\mathcal{J}_\theta d\theta) = 0$, no circulation of a real power flow along contour $L$ is presumed.

For irrotational vector field $\vec{\mathcal{J}}_\perp(\mathcal{J}_r, \mathcal{J}_\theta)$, the 2D divergence theorem looks as

$$\iint_S \vec{\nabla}_\perp \cdot \vec{\mathcal{J}}_\perp ds = \oint_L \vec{\mathcal{J}}_\perp \cdot \vec{n}dl, \tag{59}$$

where $\vec{n}$ is the outwardly directed in-plane unit normal to the contour $L$. Following the known procedure [3], we represent the power flow vector as $\vec{\mathcal{J}}_\perp = \vartheta\vec{\nabla}_\perp\upsilon$, where $\vartheta$ and $\upsilon$ are arbitrary scalar fields. With such a representation, we have $\vec{\nabla}_\perp \cdot (\vartheta\vec{\nabla}_\perp\upsilon) = \vartheta\nabla_\perp^2\upsilon + \vec{\nabla}_\perp\vartheta \cdot \vec{\nabla}_\perp\upsilon$ and $\vartheta\vec{\nabla}_\perp\upsilon \cdot \vec{n} = \vartheta\frac{\partial\upsilon}{\partial r}$. Based on these expressions, we successively consider two cases: (a) $\vartheta = \tilde{\eta}$, $\upsilon = \tilde{\eta}^*$ and (b) $\vartheta = \tilde{\eta}^*, \upsilon = \tilde{\eta}$. After simple manipulation we obtain

$$\vec{\nabla}_\perp \cdot \left(\tilde{\eta}\vec{\nabla}_\perp\tilde{\eta}^* - \tilde{\eta}^*\vec{\nabla}_\perp\tilde{\eta}\right) = \tilde{\eta}\vec{\nabla}_\perp^2\tilde{\eta}^* - \tilde{\eta}^*\vec{\nabla}_\perp^2\tilde{\eta}. \tag{60}$$

As a result, we have a flux-form Green's theorem for real power flows:

$$\iint_S \left( \tilde{\eta} \vec{\nabla}_\perp^2 \tilde{\eta}^* - \tilde{\eta}^* \vec{\nabla}_\perp^2 \tilde{\eta} \right) ds = \oint_L \left( \tilde{\eta} \frac{\partial \tilde{\eta}^*}{\partial r} - \tilde{\eta}^* \frac{\partial \tilde{\eta}}{\partial r} \right) dl . \tag{61}$$

For a given mode $n$, the homogenous boundary conditions on internal circles and, finally, the Neumann-Dirichlet boundary conditions (37), (38) on a circle $\mathcal{L} = 2\pi\mathcal{R}$ give the 2D divergence of a real-power-flow vector field $\vec{\mathcal{J}}_\perp$ equal to zero. For two eigenvalues $E_n, E_{n'}$ and, correspondingly, for two eigenfunctions $\tilde{\eta}_n, \tilde{\eta}_{n'}$ one has the orthogonality relation (40) and thus completeness of eigenfunctions $\tilde{\eta}$. In our case of MS-wave processes in a ferrite disk, we have a possibility to formulate the energy eigenstate boundary problem when real power flows, expressed by scalar wave functions $\tilde{\eta}$, propagate along the radial coordinate. As we noted above, for the signlevalued function $\tilde{\eta}$ circulation of gradient $\vec{\nabla}_\theta \tilde{\eta}$ along contour $L = 2\pi r$ is equal to zero. So, no real power flows along the azimuth coordinate are assumed for such a signlevalued function. At the same time, a reactive energy is related to the non-propagating MS fields and the reactive power flux for membrane function $\tilde{\eta}$, if any, should appear in the transverse direction, that is along the azimuth coordinate.

## V. THE MAGNETOELECTRIC EFFECT

Taking into account both the real and imaginary parts of power flows we can see that the boundary conditions for complex vector $\vec{\mathcal{J}}_\perp$ are different from the Neumann-Dirichlet boundary conditions (37), (38). On a circle $\mathcal{L} = 2\pi\mathcal{R}$ we have to have continuity of membrane function $\tilde{\eta}_n$ and a radial component of the magnetic flux density $(B_r)_n = \mu_0 \left( \mu \frac{\partial \tilde{\eta}_n}{\partial r} + \mu_a \frac{\partial \tilde{\eta}_n}{\partial \theta} \right)$. These are the electrodynamic boundary conditions expressed as:

$$\left( \tilde{\eta}_n \right)_{r=\mathcal{R}^-} - \left( \tilde{\eta}_n \right)_{r=\mathcal{R}^+} = 0 \tag{62}$$

and

$$\left( \mu \frac{\partial \tilde{\eta}_n}{\partial r} + i\mu_a \frac{1}{r} \frac{\partial \tilde{\eta}_n}{\partial \theta} \right)_{r=\mathcal{R}^-} - \left( \frac{\partial \tilde{\eta}_n}{\partial r} \right)_{r=\mathcal{R}^+} = 0 , \tag{63}$$

With such boundary conditions it becomes evident the membrane function must be not only continuous and differentiable with respect to the normal to the lateral surface, but, because of the presence of a gyrotropy term, be also differentiable with respect to a tangent to the boundary surface. The boundary conditions (37) and (38) are the so-called *essential boundary conditions*. When such boundary conditions are used, the MS-potential eigenfunctions of a differential operator form a complete basis in an energy functional space. The boundary conditions (62) and (63) are the so-called *natural boundary conditions* [19].

To restore the Neumann-Dirichlet boundary conditions (37), (38), and thus completeness of eigenfunctions $\tilde{\eta}$, we need introducing a certain surface magnetic current $j_s^{(m)}$ circulating on a lateral surface of the disk. This current should compensate the term $\left( i\mu_a \frac{1}{r} \frac{\partial \tilde{\eta}_n}{\partial \theta} \right)_{r=\mathcal{R}^-}$ in Eq. (63).

One can see that for a given direction of a bias magnetic field (that is, for a given sign of $\mu_a$), there are two, clockwise and counterclockwise, quantities of a circulating magnetic current. The current $j_s^{(m)}$ is defined by the velocity of an irrotational border flow. This flow is observable via the circulation integral of the gradient [39, 40, 42]

$$\vec{\nabla}_\theta \delta = \frac{1}{\mathcal{R}} \left( \frac{\partial \delta_\pm}{\partial \theta} \right)_{r=\mathcal{R}} \vec{e}_\theta, \qquad (64)$$

where $\delta_\pm$ is a double-valued edge wave function on contour $\mathcal{L} = 2\pi\mathcal{R}$.

On a lateral surface of a quasi-2D ferrite disk, one can distinguish two different functions $\delta_\pm$, which are the counterclockwise and clockwise rotating-wave edge functions with respect to a membrane function $\tilde{\eta}_n$. The spin-half wave-function $\delta_\pm$ changes its sign when the regular-coordinate angle $\theta$ is rotated by $2\pi$. As a result, one has the eigenstate spectrum of MS oscillations with topological phases accumulated by the edge wave function $\delta$. A circulation of gradient $\vec{\nabla}_\theta \delta$ along contour $\mathcal{L} = 2\pi\mathcal{R}$ gives a non-zero quantity when an azimuth number is a quantity divisible by $\frac{1}{2}$. A line integral around a singular contour $\mathcal{L}$: $\frac{1}{\mathfrak{R}} \oint_\mathcal{L} \left( i \frac{\partial \delta_\pm}{\partial \theta} \right) (\delta_\pm)^* d\mathcal{L} = \int_0^{2\pi} \left[ \left( i \frac{\partial \delta_\pm}{\partial \theta} \right) (\delta_\pm)^* \right]_{r=\mathfrak{R}} d\theta$ is an observable quantity. Because of the existing the geometrical phase factor on a lateral boundary of a ferrite disk, MS oscillations are characterized by a pseudo-electric field (the gauge field) $\vec{\epsilon}$. The pseudo-electric field $\vec{\epsilon}$ can be found as $\vec{\epsilon}_\pm = -\vec{\nabla} \times (\vec{\Lambda}_\epsilon^{(m)})_\pm$. The field $\vec{\epsilon}$ is the Berry curvature. The corresponding flux of the gauge field $\vec{\epsilon}$ through a circle of radius $\mathfrak{R}$ is obtained as: $K \int_S (\vec{\epsilon})_\pm \cdot d\vec{S} = K \oint_\mathcal{L} (\vec{\Lambda}_\epsilon^{(m)})_\pm \cdot d\vec{\mathcal{L}} = K (\Xi^{(e)})_\pm = 2\pi q_\pm$, where $(\Xi^{(e)})_\pm$ are quantized fluxes of pseudo-electric fields, $K$ is the normalization coefficient. Each MS mode is quantized to a quantum of an emergent electric flux. There are the positive and negative eigenfluxes. These different-sign fluxes should be nonequivalent to avoid the cancellation. It is evident that while integration of the Berry curvature over the regular-coordinate angle $\theta$ is quantized in units of $2\pi$, integration over the spin-coordinate angle $\theta'$ $\left(\theta' = \frac{1}{2}\theta\right)$ is quantized in units of $\pi$. The physical meaning of coefficient $K$ concerns the property of a flux of a pseudo-electric field. The Berry mechanism provides a microscopic basis for the surface magnetic current at the interface between gyrotropic and nongyrotropic media. Following the spectrum analysis of MS modes in a quasi-2D ferrite disk one obtains pseudo-scalar axion-like fields and edge chiral magnetic currents [39]. The anapole moment for every mode $n$ is calculated as [39, 40, 42]:

$$a_\pm^{(e)} \propto \mathcal{R} \int_0^d \oint_\mathcal{L} \left[ \vec{j}_s^{(m)}(z) \right]_\theta \cdot d\vec{l} dz. \qquad (65)$$

It follows that different orientations of an electric moment $a_\pm^{(e)}$ (parallel or antiparallel with respect to $H_0$) correspond to different energy levels. The energy splitting between two cases, $a_\pm^{(e)} H_0 > 0$ and $a_\pm^{(e)} H_0 < 0$, is defined as the magnetoelectric (ME) energy [39, 40, 42]. The ME energy is the potential energy of a MS-mode ferrite disk in an external magnetic field. It is alike to another type of the potential energy of a magnetised body in an external magnetic field – the Zeeman energy.

With taking into account the ME energy, the MS-potential membrane wave function is expressed as: $(\tilde{\varphi}_\pm)_{r=\Re^-} = \delta_\pm (\tilde{\eta})_{r=\Re^-}$. For any mode $n$, the function $\tilde{\varphi}_n$ is a two-component sprinor pictorially denoted by two arrows:

$$\tilde{\varphi}_n^{\uparrow\downarrow}(\vec{r}, \theta) = \begin{bmatrix} \tilde{\varphi}_n^\uparrow \\ \tilde{\varphi}_n^\downarrow \end{bmatrix} = \tilde{\eta}_n(\vec{r}, \theta) \begin{bmatrix} e^{-\frac{1}{2}i\theta} \\ e^{+\frac{1}{2}i\theta} \end{bmatrix} \tag{66}$$

Circulation of gradient $\vec{\nabla}_\theta \tilde{\varphi}$ along contour $L = 2\pi r$ is not equal to zero. So, we can observe the angular momentum due to the power-flow circulation:

$$\vec{L}_z = \oint_L \vec{r} \times \frac{1}{r} \left( \vec{\mathcal{J}}_\theta^{\uparrow\downarrow} \right)_n dl, \tag{67}$$

where

$$\left( \vec{\mathcal{J}}_\theta^{\uparrow\downarrow} \right)_n = \frac{1}{4} \omega |A_n|^2 |\xi_n(z)|^2 \mu_0 \mu_a \frac{1}{r} \left[ \tilde{\varphi}_n \left( \frac{\partial \tilde{\varphi}_n}{\partial \theta} \right)^* - \tilde{\varphi}_n^* \frac{\partial \tilde{\varphi}_n}{\partial \theta} \right] \vec{e}_\theta. \tag{68}$$

The angular momentum $\vec{L}_z$ is an intrinsic property of a ferrite-disk particle, unrelated to any sort of motion in space. The direction of rotation is correlated with the direction of an anapole moment with repect of a bias magnetic field. The rotational energy should be equal to the ME energy. The condition

$$E_{rotational} = E_{ME} \tag{69}$$

is the necessary conditions to get a Hermitian Hamiltonial.

With use of Eq. (57) we have now

$$\left( \vec{\mathcal{J}}_z \right)_n = -\frac{i\omega}{4} \mu_0 |A_n|^2 |\tilde{\eta}(r,\theta)|^2 \left[ \xi_n \left( \frac{\partial \xi_n}{\partial z} \right)^* - \xi_n^* \frac{\partial \xi_n}{\partial z} \right] \vec{e}_z + \oint_L \vec{r} \times \frac{1}{r} \left( \vec{\mathcal{J}}_\theta^{\uparrow\downarrow} \right)_n dl. \tag{70}$$

The first term in this momentum density is associated with the translational motion, whereas the second term is associated with circulated flow of energy in the rest frame of the disk. For such a joint effect of the translational and circulating motions, the mebrane function $\tilde{\varphi}_n$ should be represented as a four-component sprinors described by a Hermitian adjoint operator.

## VI. CONCLUSION

Electromagnetic responses of electrostatic plasmon resonances in optics and magnetostatic magnon resonances in microwaves give rise to a strong enhancement of local fields near the surfaces of subwavelength particles. Uncertainty in definition of the electric or magnetic field components in the near-field regions of subwavelength particles raises the question of energy eigenstates of quasistatic oscillations. In this paper, we argue that staying in frames of the quasistatic description of oscillations in a subwavelength particle, such an energy eigenstate problem can be properly formulated based on scalar-potential wave functions. In this case, one should observe quasistatic-wave retardation effects caused by long range dipole-dipole correlations. We showed that in a case of MS-potential scalar wave function $\psi$ one can observe such quasistatic retardation effects and obtain a proper formulation of the energy eigenstate problem based on the Schrödinger-like equation.

In literature, it is argued that the electron spin may be regarded as an angular momentum generated by a circulating flow of energy in the wave field of the electron [43, 44]. In this paper, we showed that in a subwavelength ferrite-disk particle one can observe the angular momentum due to the power-flow circulation of double-valued edge MS wave functions. For incident electromagnetic wave, this magnon subwavelength particle emerges as a singular point carrying quanta of angular momentum.